# Deformation-induced amorphous complexion transitions elevate strength and ductility

Masoud Ahmadi[1,*], Jin Qin[2], Gabrielle Tiphéne[1,3], Mohamed Charai[4], Alejandro Gómez-Pérez[5], Khalid Hoummada[4], Thomas Pardoen[1,6], Matteo Ghidelli[2], Hosni Idrissi[1]

[1] Institute of Mechanics, Materials and Civil Engineering (iMMC), UCLouvain, Louvain-la-Neuve, B-1348, Belgium

[2] Laboratoire des Sciences des Procédés et des Matériaux (LSPM)-CNRS, UPR 3407, Université Sorbonne Paris Nord, 93430 Villetaneuse, France

[3] PSL Research University, Chimie ParisTech, Institut de Recherche de Chimie Paris, Paris, France

[4] Aix-Marseille Université, CNRS, IM2NP, 13397, Marseille, France

[5] NanoMegas SPRL, Rue Émile Claus 49 bte 9, 1050 Brussels, Belgium

[6] WEL Research Institute, Avenue Pasteur 6, 1300, Wavre, Belgium

* Corresponding author: masoud.ahmadi@uclouvain.be

https://orcid.org/0000-0002-5367-5621

**Abstract**

Grain boundary engineering is a major avenue for tailoring the mechanical behavior of polycrystalline materials. Grain boundary complexions, including amorphous intergranular films, are classically accessed through thermal driving forces and solute segregation. Here, we discover that plastic deformation can drive amorphous complexion transitions at room temperature in a chemically primed nanocrystalline binary CuZr alloy. High-resolution and four-dimensional scanning transmission electron microscopy reveal that the amorphous complexions preferentially emerge at incoherent twin boundaries. Spatially-resolved electron pair distribution function analysis at the atomic scale — the local-order characterization of amorphous complexions — demonstrates short-range and medium-range order gradients from crystal-templated interfaces to a metallic-glass-like core. We thus uncover a novel *amorphous complexion transformation-induced plasticity* mechanism that concurrently increases the yield strength, fracture strain, and tensile toughness about a factor of two relative to a designed reference material. Our findings establish mechanical deformation as a non-thermal pathway to trigger amorphous interfacial states for enhancing damage tolerance in nanostructured metals.

## Main

Nanocrystalline materials exhibit remarkable strength due to high density of grain boundaries (GBs). These interfaces govern the resistance to failure, as grain sizes on the order of tens of nanometers localize deformation within the interconnected GB network, thereby compromising damage tolerance, ductility and toughness. Engineering the GB character, rather than grain size alone, offers a fundamentally different route for overcoming the strength–ductility trade-off in nanocrystalline metals[1–6].

Grain boundary complexions[7] — interfacial phases governed by thermodynamic variables including temperature, chemical potential, and pressure — provide a powerful framework to implement this approach[8,9]. Cantwell et al.[10,11] defined complexion transitions as first-order structural transformations of the intergranular region, and Frolov and Mishin[12] established a thermodynamic framework in which GB free energy depends simultaneously on temperature, composition, and stress state, providing theoretical basis for complexion transitions. Among the complexion categories introduced by Dillon-Harmer[7], amorphous intergranular films (AIFs) represent a high-disorder state. These nanoscale amorphous boundary films act as effective dislocation sinks as shown extensively by Turlo and Rupert[13]. CuZr is a favorable system for thermally stabilized AIFs formation owing to Zr's negative enthalpy of mixing, glass forming ability and its large atomic radius mismatch with the Cu matrix[14]. Thermally stabilized AIFs in CuZr can absorb multiple dislocations, delay intergranular void formation, and increase GB damage tolerance[15]. Subsequent atomistic and experimental studies demonstrated that nanocrystalline CuZr alloys containing AIFs involve larger fracture strain, with the AIF boundary network suppressing strain localization and promoting enhanced toughness[16]. In tensile experiments, thermally stabilized AIF-containing CuZr alloys exhibit retarded failure and reduced shear-dominated fracture[17].

However, AIF formation has remained exclusively in the domain of thermodynamic control through elevated-temperature annealing to drive solute segregation and boundary structural transitions[11,18]. Stress-induced amorphization reported previously is confined to shear-band interiors or transient bulk phenomena (predominantly in multicomponent non-binary alloys)[19–26], without exhibiting the characteristic morphology, spatial distribution, chemistry, and intergranular stability of GB AIF complexions.

Here, we discover that plastic deformation drives AIF complexion transitions in a binary nanocrystalline CuZr alloy at room temperature, without thermal activation. Through a combination of automated crystal orientation mapping in TEM (ACOM-TEM), in-situ TEM push-to-pull (PTP) nanomechanical testing, atom probe tomography (APT), four-dimensional scanning transmission electron microscopy (4D-STEM), and spatially-resolved electron pair distribution function (e-PDF) analysis, we demonstrate the unexpected emergence of nanoscale AIF complexions, and provide the first atomic-scale structural characterization of deformation-induced amorphous complexions revealing short-range and medium-range order gradients. We unveil, for the first time, the GB mechanism of *amorphous complexion transformation-induced plasticity (AC-TRIP)*, which simultaneously boosts strength, ductility and tensile toughness. Overall, our results demonstrate plastic deformation as a fundamentally distinct, non-thermal strategy to harness amorphous GB complexion states, offering promising implications for engineering damage-tolerant nanostructured metals.

**Nanostructure design through grain boundary engineering**

Two CuZr alloy thin film systems were designed and produced. The primary sample, CuZr-1, contains 6.2 at.% Zr, whereas the reference sample, CuZr-2, contains 2.8 at.% Zr. We first characterize the microstructure and chemical composition of both films in the as-deposited state. Fig.1 summarizes the ACOM-TEM, HRTEM, and APT analyses of the two alloys.

The CuZr-1 film (Fig. 1a) exhibits columnar grains along the out-of-plane direction and a distinctive bimodal microstructure. The columnar grains contain a high density of twins with curved interfaces. In addition, narrower columns without detectable twins are also present. Two characteristic grain sizes can be distinguished: (i) the columnar grain width and (ii) the average twin thickness. Accordingly, CuZr-1 exhibits an average columnar grain width of 54 nm, while the twins within the wider columns have an average thickness of ~10 nm (see Methods). As shown in Fig. 1b, Σ3 twin boundaries account for approximately 52% of all grain boundaries in the CuZr-1 film (see Fig. S1 for the corresponding misorientation profile). Fig. 1c further reveals the nature of these twins. The representative HRTEM image shows an incoherent twin boundary (ITB) associated with a local 9R structure[27,28]. The corresponding fast Fourier transforms (FFTs) confirm the presence of the 9R structure through the appearance of additional diffraction spots[28].

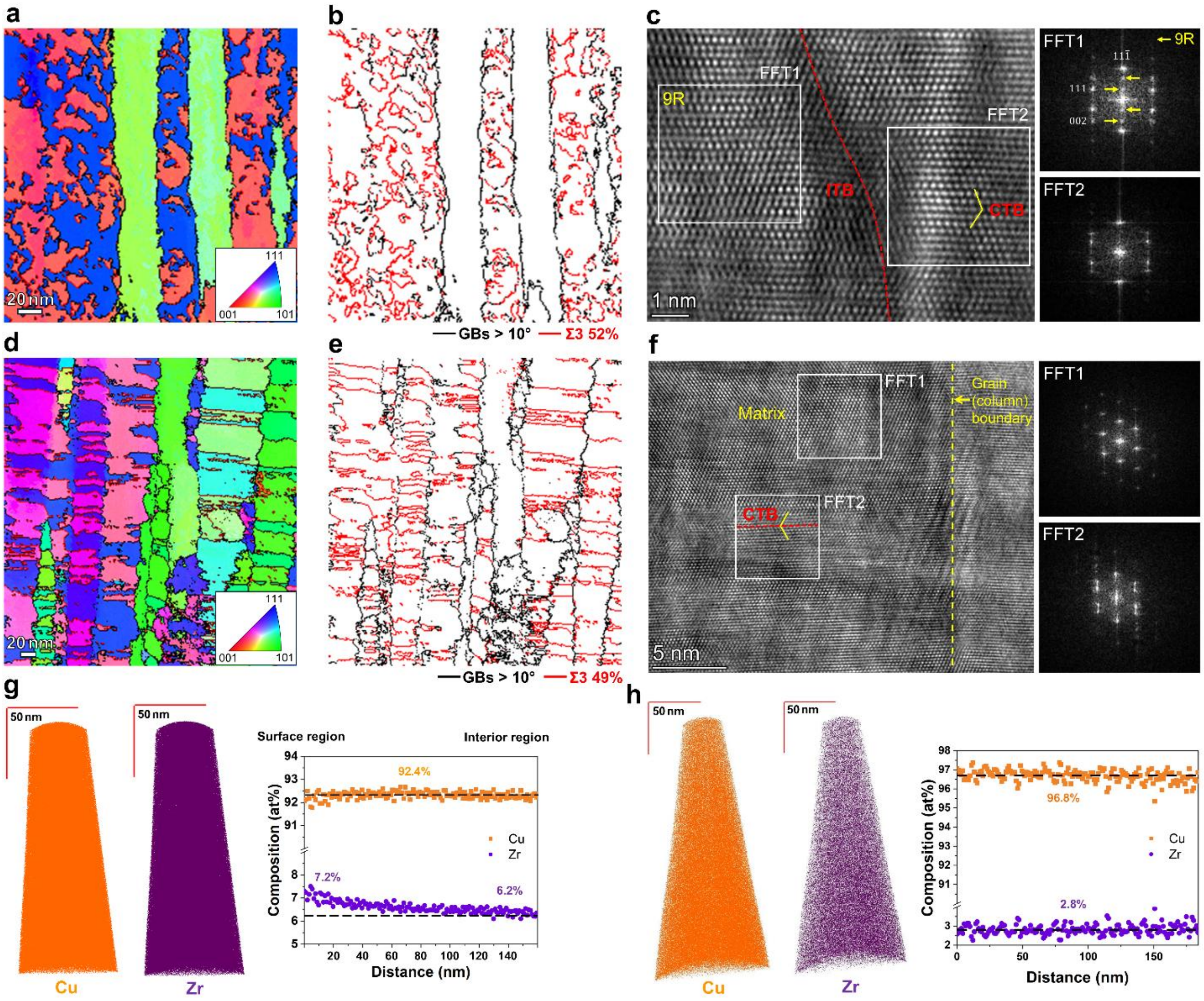

**Fig. 1: Nanostructure and chemical composition of the as-deposited CuZr alloys prior to deformation.**

**a,** ACOM-TEM orientation map of CuZr-1 according to the inverse pole figure (IPF) referenced to the film growth direction (vertical). **b,** ACOM-TEM map highlighting grain and twin boundaries in CuZr-1. **c,** Representative HRTEM image and corresponding FFTs showing ITBs and the associated 9R structure in CuZr-1. **d,** ACOM-TEM orientation map of CuZr-2 according to the IPF referenced to the film growth direction. **e,** ACOM-TEM map illustrating grain and twin boundaries in CuZr-2. **f,** Representative HRTEM image and corresponding FFTs showing CTBs and a columnar boundary in CuZr-2. **g,** APT reconstruction demonstrating the spatial distribution/content of Cu and Zr in CuZr-1. **h,** APT reconstruction showing the spatial distribution/content of Cu and Zr in CuZr-2.

Fig. 1d-f present ACOM-TEM and HRTEM micrographs of the CuZr-2 film. Although this alloy also exhibits columnar growth, the microstructure differs substantially from that of CuZr-1. The average columnar grain width is 63 nm, and most grains contain abundant twins with an average thickness of ~11 nm oriented perpendicular to the growth direction (see Fig. S2). The corresponding misorientation profile is shown in Fig. S1. The HRTEM image in Fig. 1f shows that these twins are bounded by atomically sharp interfaces parallel to {111} plane, characteristic of Σ3 {111} coherent twin boundaries (CTBs)[29]. Thus, the primary structural distinction of CuZr-1 and CuZr-2 lies in the character of the twin boundaries (i.e. ITBs in CuZr-1 versus CTBs in CuZr-2) rather than the characteristic length scales. No amorphous regions were detected in either film in the as-deposited state prior to deformation. This observation is consistent with room-temperature magnetron sputtering deposition, for which the limited atomic mobility is insufficient to promote Zr segregation to grain boundaries and the subsequent formation of thermally stabilized AIFs. Bright-field TEM (BF-TEM) images and X-ray diffraction (XRD) results are given in Fig. S2-S3.

The APT results shown in Figs. 1g-h further highlight the nominal Zr content of 6.2 at.% (with ~1 at.% Zr rise at the surface) and 2.8 at.% for CuZr-1 and CuZr-2, respectively. Both Cu and Zr are distributed uniformly throughout the volumes, indicating that Zr remains in solid solution without any detectable precipitation and segregation to specific features like grain boundaries. Although Zr exhibits low equilibrium solubility in Cu, the non-equilibrium conditions of magnetron sputtering retain the metastable CuZr-1 alloy in a supersaturated state. This conclusion is further supported by APT analysis on a large (~250 nm) specimen performed perpendicular to the growth direction (spanning several GBs), see supplementary Fig. S4-S5 and Note 1 for details.

These results establish two chemically homogeneous CuZr alloys with distinct GB structures but no pre-existing AIF. CuZr-1 combines a supersaturated solid solution with a high fraction of ITBs and local 9R structures, whereas CuZr-2 consists predominantly of CTBs. This well-defined microstructural contrast provides the basis for isolating the role of GB character in this work.

**Deformation-induced amorphous complexion transitions**

We now examine the post-deformation nanostructures of CuZr-1 after nanoindentation and in-situ TEM PTP tensile tests (see Methods for descriptions). Fig. 2 presents a BF-TEM image of a FIB cross-section prepared beneath a nanoindentation imprint in CuZr-1, together with the ACOM-TEM maps. Notably, in the immediate vicinity of the indent contact area, severe deformation has produced refined grains. The orientation map further indicates that regions formerly occupied by twins have undergone slight orientation changes, consistent with twin boundary distortion or grain rotation. The most important finding, however, is the formation of isolated amorphous patches within the high-stress zone, as revealed by the ACOM-TEM phase map in Fig. 2d. These amorphous regions are spatially discontinuous, distributed within the highly stressed zone, and range from ~3 to 20 nm in size.

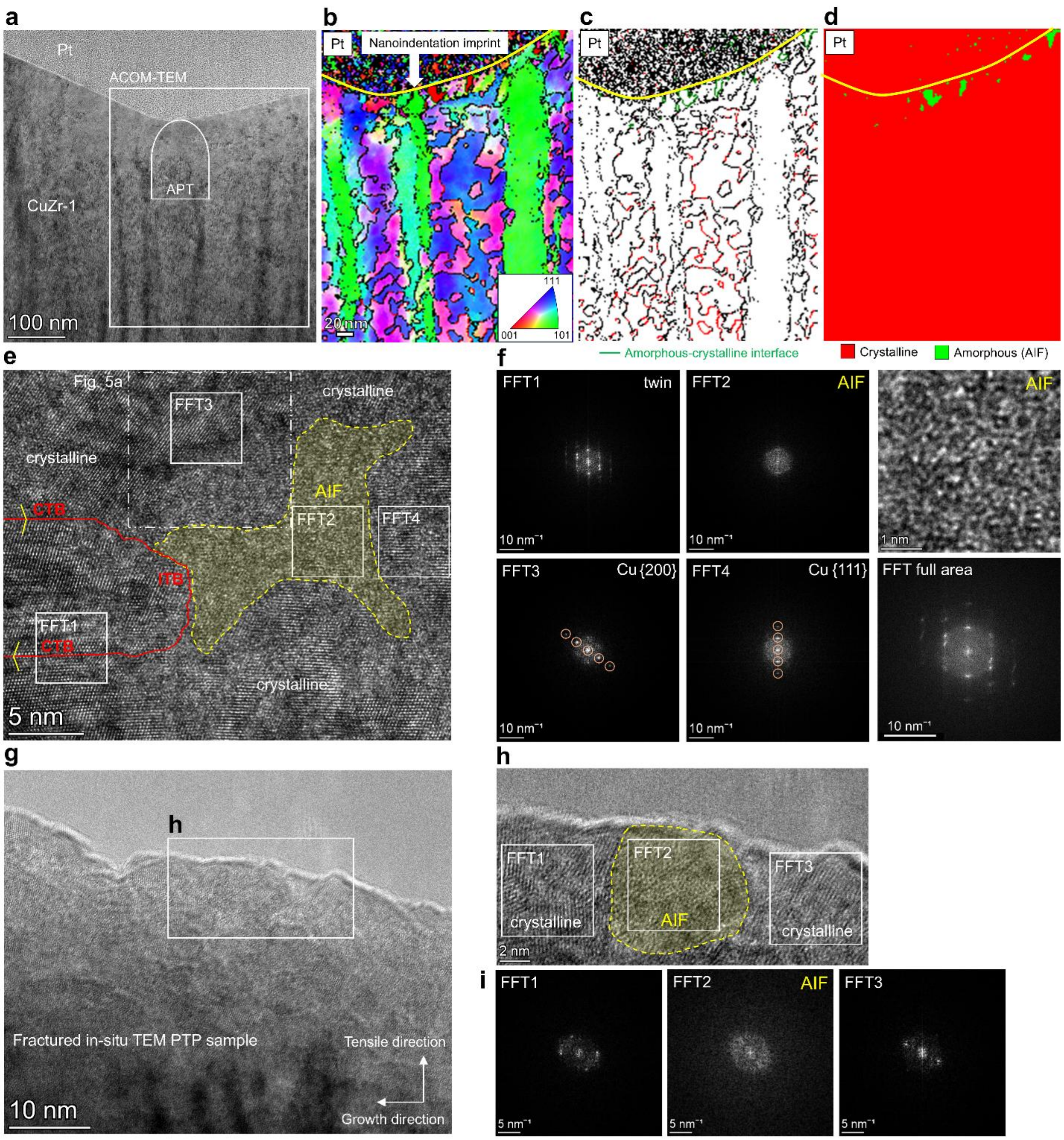

**Fig. 2: Post-deformation nanostructure of CuZr-1 revealing deformation-driven AIF complexion transitions.**

**a,** BF-TEM image of the FIB lamella showing the nanoindentation imprint and the regions selected for post-deformation ACOM-TEM and APT analyses. **b,** Post-deformation ACOM-TEM orientation map according to the IPF referenced to the film growth direction. **c,** Post-deformation ACOM-TEM map highlighting grain and twin boundaries. **d,** Post-deformation ACOM-TEM phase map showing the coexistence of crystalline and amorphous regions. **e,f,** Representative HRTEM image and corresponding FFTs demonstrating the coexistence of crystalline grains and the deformation-induced AIF complexion formed at the ITB after nanoindentation. **g,** Fractured edge of the in-situ TEM PTP tensile specimen. **h,** HRTEM image acquired from the immediate vicinity of the fractured edge, showing the formation of AIF complexions after tensile deformation. **i,** Corresponding FFTs confirming the amorphous character of the AIF complexions.

Fig. 2e-f show a representative amorphous domain confined between adjacent highly deformed crystalline grains. The FFT of the amorphous region displays a diffuse halo with no discrete diffraction spots, whereas the surrounding crystalline regions exhibit {111} and {200} reflections. The identified AIF formed at the ITB and exhibits a size exceeding 15 nm, larger than those reported for thermally stabilized AIFs[15], (see additional results in Fig. S6).

AIF formation is not restricted to nanoindentation (with compressive loading). Fig. 2g-i present HRTEM micrographs and FFTs from the fracture surface of the in-situ tensile specimen. Nanoscale amorphous domains are found at the fracture edge, bounded by crystalline grains exhibiting distinct FFT reflections (see Fig. S7 for supporting evidence). This AIF was also most likely formed at ITB; however, confirming its exact nucleation site was not possible due to the limitations inherent to PTP experiment. ACOM-TEM analysis (Fig. S8) of the fractured edge gives further insight into these amorphous states. The fully crystalline CuZr-1 film has therefore undergone an AIF complexion transition under both compressive and tensile deformation modes. The CuZr-2 reference film was investigated under identical nanoindentation and in-situ TEM tensile conditions; no AIF formation was detected in any case (Fig. S9-S10), confirming that the deformation-induced AIF transition is unique to CuZr-1. This will be elaborated in the final section.

**Atomic-scale characterization of structural order in amorphous complexions**

Since thermally stabilized AIFs are typically only 1 to 5 nm thick[30], characterizing their short- and medium-range order experimentally has been challenging. Owing to the larger deformation-induced AIFs found in this study, it was possible to experimentally characterize the local atomic order gradient across the entire AIF for the first time.

Fig. 3a presents a 4D-STEM e-PDF map of an AIF region encompassing three distinct zones: the nanocrystalline grain interior, the crystal-AIF interface and the AIF core. Along a line scan from the crystalline grain (position 1) through the interface to the core (position 10), the first structure factor S(Q) peak near $Q \approx 3$ Å$^{-1}$ broadens progressively—sharpest at the crystalline reference and broadest at the AIF core—indicating a systematic loss of structural order (Fig. 3b). Further, Fig. 3c-d display the reduced pair distribution function G(r) profile across a deformation-induced AIF, revealing the atomic-scale structural transition from crystalline to amorphous. The Cu-Cu first-coordination-shell distance ($r \approx 2.5$ Å) is preserved throughout — consistent with the metallic nearest-neighbor distance expected for a Cu-rich amorphous phase[31] — while structural order is systematically lost with distance from the crystalline grain. The first G(r) peak for the AIF region (Fig. 3d) has slightly shifted rightward compared to the crystalline area, possibly due to volume change associated with amorphization. Notably, the single second-coordination peak of the FCC crystalline reference splits into two resolved sub-peaks (between $r \approx 3.5$ Å and $r \approx 4.5$ Å) in the AIF core. This peak splitting is the characteristic signature of the transition from long-range order to metallic glass short/medium-range order[32]: the uniform second-shell distance distribution of FCC (a single peak) transforms into the bimodal second-shell distribution characteristic of icosahedral and polyhedral clusters in metallic glasses[31,33]. The progressive development of the split — evident as a broadening and shoulder in curves 5-6 (at the interface) and fully resolved double peaks in curves 9-10 (AIF core) — captures the spatial scale over which FCC coordination transforms to metallic glass coordination geometry. These observations provide direct experimental support for the short-range order gradients in AIFs predicted by molecular dynamics simulation[34] and establish the deformation-AIF as a structurally distinct complexion with internal spatial order.

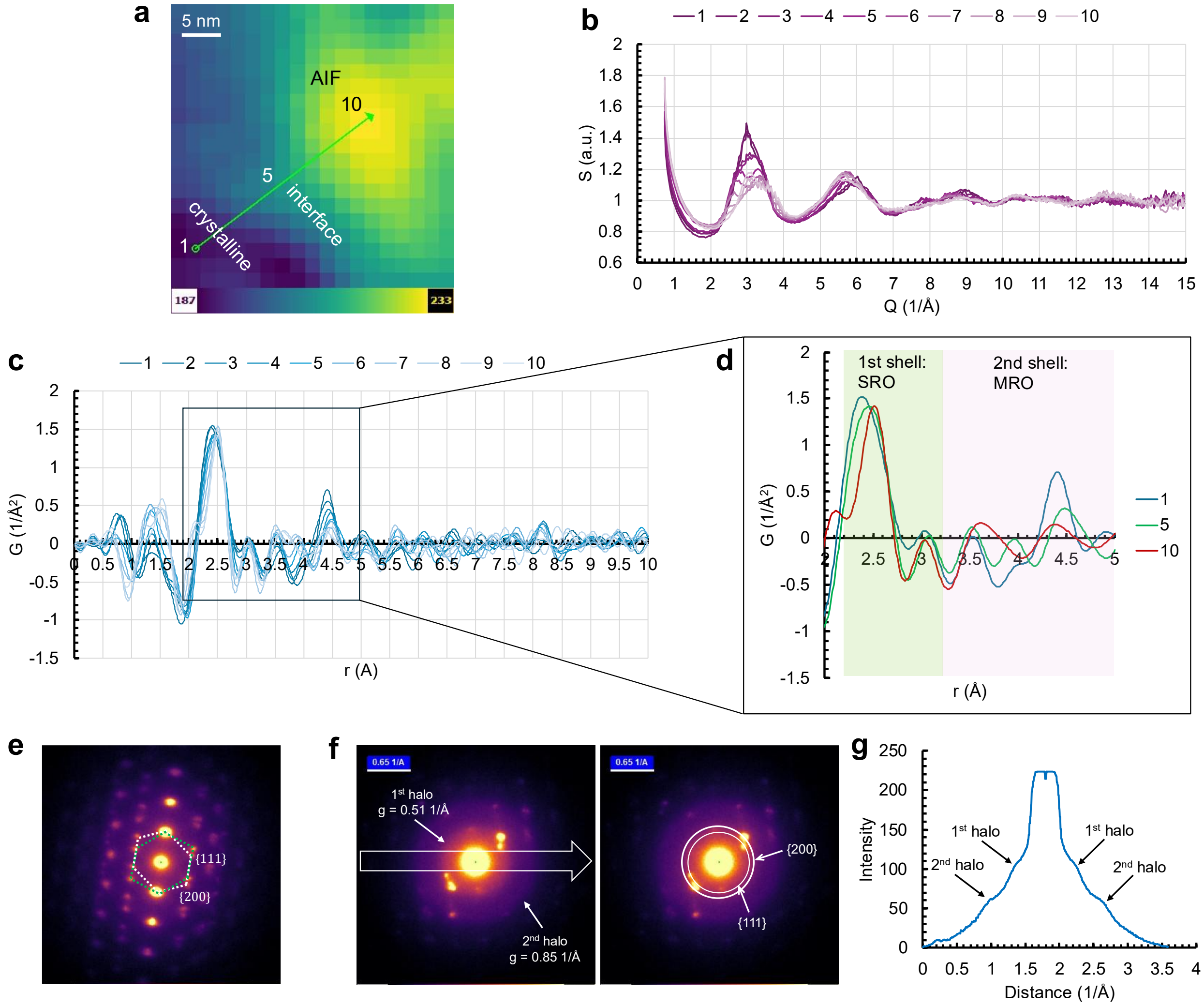

**Fig. 3: Characterization of the structural order within the AIF complexion via 4D-STEM and e-PDF analyses.**

**a**, 4D-STEM e-PDF map of the region of interest showing the deformation-induced AIF complexion, its interface with the adjacent crystalline grains, and the fully crystalline region. **b**, Spatially resolved ePDF S(Q) profiles acquired along the line scan shown in **(a)** from point 1 (crystalline) to point 10 (AIF core). **c**, Spatially resolved ePDF G(r) curves across the same line profile, Curves 1–10 progress from the crystalline grain interior (curve 1, dark blue) to the AIF core (curve 10, light blue), with measurements collected at approximately equal spatial intervals, revealing the gradual evolution from crystalline to metallic-glass-like character. **d**, Enlarged view of the ePDF G(r) curves comparing point 1 (crystalline, blue), point 5 (interface, green) and point 10 (AIF core, red). **e**, 4D-STEM nanobeam electron diffraction pattern acquired under the indent away from the transition zone. **f,g**, 4D-STEM nanobeam electron diffraction pattern acquired in the AIF core (point 10) and the corresponding radial and intensity analyses showing the $1^{st}$ and $2^{nd}$ diffuse halos. The intensity profile in (**g**) is along the thick white arrow in (**f**). The probe size in these experiments was ~2 nm.

Nanobeam diffraction patterns from the 4D-STEM datasets corroborate these findings. The pattern from a crystalline region away from the transition zone (Fig. 3e) indexes to the {111} and {200} reflections of the twins, whereas the pattern from the AIF core (position 10; Fig. 3f) exhibits two diffuse halos characteristic of a metallic amorphous phase. The first halo (at g ≈ 0.51 1/Å) coincides with a few bright residual spots; measuring d-spacings for two pairs of spots yields $d_1 \approx 2.17$ Å and $d_2 \approx 1.85$ Å, consistent with the {111} (d = 2.09 Å) and {200} (d = 1.81 Å) d-spacings of FCC Cu. The amorphous halo at the AIF core thus coincides with the twin reflections, supporting that AIFs form at twin boundaries. Correspondingly, in the S(Q) representation (Fig. 3b), first peak at Q ≈ 3.0-3.5 $Å^{-1}$ spans the {111} and {200} positions. This S(Q) peak and the G(r) first peak at r ≈ 2.5 Å — slightly contracted relative to pure FCC Cu (2.56 Å) — show the same nearest-neighbor correlations from reciprocal- and real-space perspectives.

These results (Fig. 3) suggest that the crystalline grains surrounding AIF impose structural order on the adjacent amorphous material: near the interface, the shared twin geometry templates the local atomic packing into dense, FCC-like arrangements, whereas the AIF core — without this constraint — shows a metallic-glass-like atomic arrangement with icosahedral and polyhedral motifs. These findings also open avenues for correlating short- and medium-range order with chemical patterning in the broader context of complexion science[35].

**Multimodal in-situ nanomechanical testing**

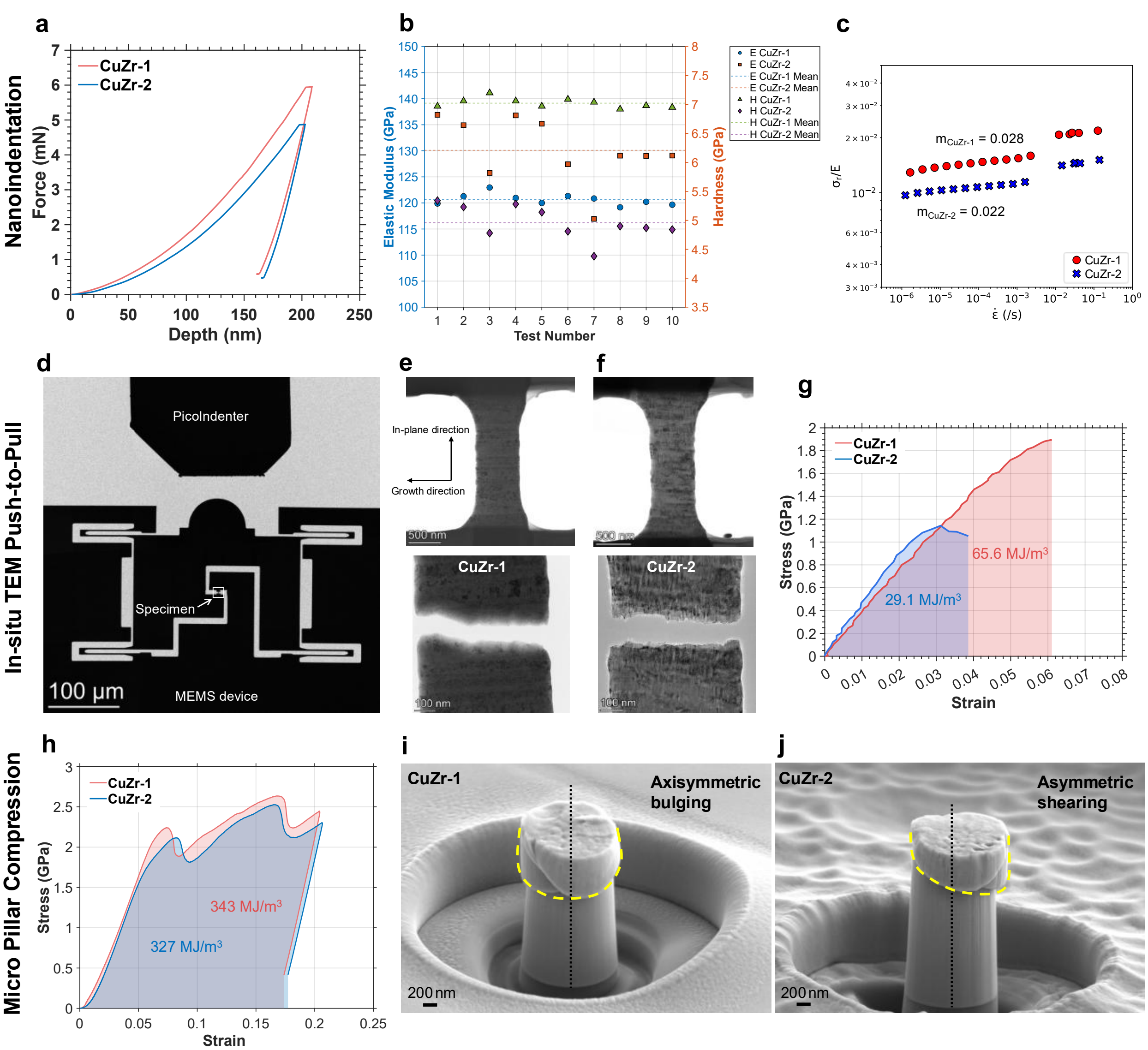

**Fig. 4: Mechanical response of CuZr-1 and CuZr-2 across multiple nano- and micro-mechanical testing modalities.**

**a,** Representative nanoindentation load–displacement response for CuZr-1 and CuZr-2. **b,** Elastic modulus and hardness of the two alloys from ten nanoindentation measurements. **c,** Nanoindentation relaxation behavior and strain-rate sensitivity of the two alloys. **d,** Overview TEM image of the in-situ TEM PTP tensile testing configuration, including the MEMS device, the specimen, and the PicoIndenter. **e,** BF-TEM images of the CuZr-1 specimen before and after in-situ TEM tensile testing. **f,** BF-TEM images of the CuZr-2 specimen before and after in-situ TEM tensile testing. **g,** Engineering stress–strain curves obtained from the in-situ TEM PTP tensile experiments. **h,** Engineering stress–strain curves from micropillar compression tests of CuZr-1 and CuZr-2. **i,** Post-mortem SEM image of the CuZr-1 micropillar after compression. **j,** Post-mortem SEM image of the CuZr-2 micropillar after compression. The yellow dashed lines indicate the borders of the pillar caps after failure, whereas the black dashed lines indicate the axis of symmetry of the pillars.

Nanoindentation results (Fig. 4a-b) show for CuZr-1 a hardness equal to 7.02 higher than 4.96 GPa obtained for CuZr-2. The elastic modulus is equal to 120.6 and 130.1 GPa for CuZr-1 and CuZr-2, respectively. The yield strength estimated from nanoindentation (see Methods) is equal to 2.23 and 1.52 GPa for CuZr-1 and CuZr-2, respectively, consistent with the H/3 relationship. Notably, to our knowledge, the 2.23 GPa yield strength obtained here is among the highest reported to date for Cu-based binary nanocrystalline alloys[36,37]. Nanoindentation relaxation measurements over strain rates spanning from $10^{-6}$ up to $10^{-1}$ (1/s) indicate a higher strain-rate sensitivity (m) for CuZr-1 ($m_{CuZr-1}$=0.028 and $m_{CuZr-2}$=0.022). In-situ TEM PTP tensile tests (Fig. 4d-g and Videos 1-2) yield a 0.2% proof stress $\sigma_{y\,(0.2\%)} =$ 1.59 GPa, maximum tensile stress $\sigma_{max} =$ 1.9 GPa, and fracture strain $\varepsilon_f =$ 0.061 for CuZr-1, compared with $\sigma_{y\,(0.2\%)} =$ 0.88 GPa, $\sigma_{max} =$ 1.14 GPa, and $\varepsilon_f =$ 0.038 for CuZr-2. Integration of the stress-strain curves gives tensile toughness values of 65.6 and 29.1 MJ $m^{-3}$ for CuZr-1 and CuZr-2, respectively. Moreover, CuZr-1 shows a sustained, pronounced strain hardening behavior until failure, while CuZr-2 exhibits post-peak softening after reaching the maximum stress. The nanoindentation and in-situ tensile results demonstrate the superior mechanical behavior of CuZr-1, which undergoes amorphous complexion transitions. Compared with CuZr-2, CuZr-1 exhibits a 1.8× higher yield strength, 1.6× larger fracture strain, and 2.3× higher tensile toughness, thereby mitigating the conventional strength-ductility trade-off.

Nanoindentation probes deformation along the film growth direction, reducing the influence of columnar grain-boundary fracture. In contrast, due to geometrical constraints during FIB preparation, the in-situ tensile specimens are made parallel to the film plane, promoting fracture along columnar boundaries (see Fig. 4e, Supplementary ACOM-TEM analysis in Fig. S8 and Note 2). Consequently, the ductility measured by PTP should be considered a lower-bound estimate of the intrinsic ductility that the material can achieve.

Representative micropillar compression results (Fig. 4h-j) corroborate the superior mechanical performance of CuZr-1 under compressive loading, yielding $\sigma_{y\,(0.2\%)} =$ 2.1 GPa, $\sigma_{max} =$ 2.63 GPa, and compressive toughness of 343 MJ $m^{-3}$, compared to $\sigma_{y\,(0.2\%)} =$ 1.8 GPa, $\sigma_{max} =$ 2.52 GPa, and compressive toughness of 327 MJ $m^{-3}$ for CuZr-2. Post-mortem SEM observations of the deformed micropillars (Fig. 4i-j) reveal a qualitative difference in the failure morphologies. The CuZr-2 pillar exhibits a pronounced, asymmetric cap with a one-sided overhang, consistent with a dominant shear-offset event traversing a large fraction of the pillar cross-section. The CuZr-1 pillar exhibits a more axisymmetric, uniformly distributed cap bulge with a smaller overhang, consistent with more distributed plastic accommodation potentially associated with the thick AIF-decorated boundary network[38]. This distinction agrees with the higher post-burst stress recovery observed for CuZr-1 in the compressive engineering stress-strain response (Fig. 4h). (More details in Supplementary Fig. S12-S13).

**Mechanism of amorphous complexion transformation-induced plasticity (AC-TRIP)**

Fig. 5a-b provide representative HRTEM and inverse FFT (IFFT) micrographs of the deformed CuZr-1 beneath the nanoindentation imprint (adjacent to AIF as indicated in Fig. 2e). Fig. 5c (and Fig. S14) shows site-specific APT results corresponding to an area in the amorphization zone beneath the nanoindentation imprint (designated in Fig. 2a).

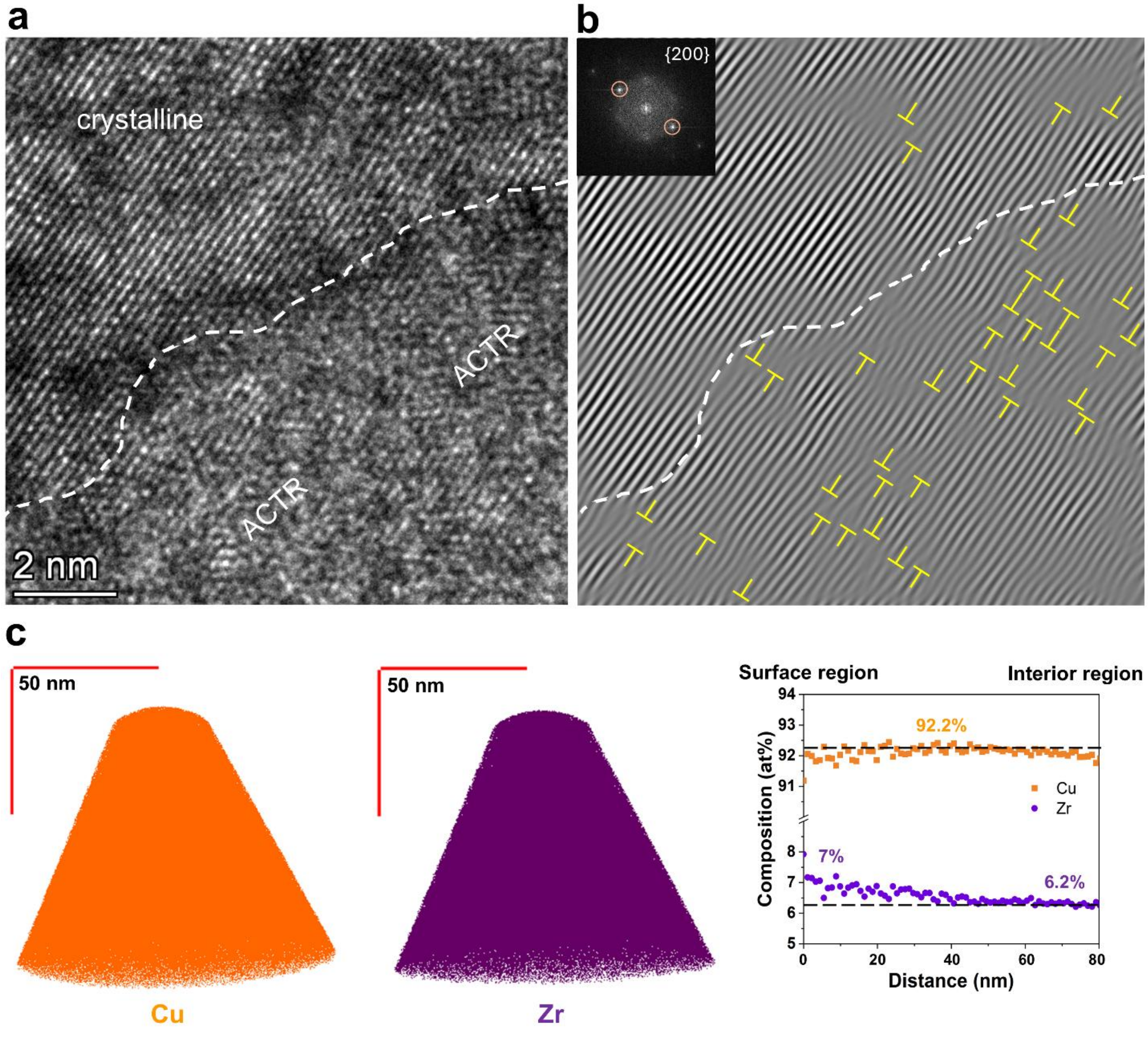

**Fig. 5: Plasticity mechanism evaluation of the primary alloy CuZr-1.**

**a,** Representative HRTEM image of the deformed CuZr-1 alloy beneath the nanoindentation imprint (adjacent to AIF as indicated in Fig. 2e). **b,** IFFT image corresponding to (**a**), revealing abundant dislocations (yellow T symbols) at the amorphous-crystalline transition region (ACTR). The {200} FFT spots for generating the IFFT micrograph are designated in the upper left corner. **c,** APT reconstruction acquired from the deformation-induced amorphous region beneath the nanoindentation imprint.

These results confirm a homogeneous distribution of Cu and Zr elements, like in the as-deposited film (see Fig. 1). Therefore, no Zr redistribution was detected after plastic deformation. The transition therefore involves no detectable long-range solute transport: plastic deformation acts as a kinetic trigger for AIF formation in a boundary network already chemically primed by supersaturated Zr (6.2 at%) in solid solution. Conversely in CuZr-2, the Zr content (2.8 at.%) appears insufficient to saturate the boundaries, leading to the absence of AIF transition.

To elaborate, the high density of dislocation cores observed at the amorphous-crystalline transition region (ACTR)[39] in Fig. 5a-b provide local insights into the deformation mechanism underlying AIF formation. In Cu-based alloys with low stacking-fault energy, ITBs are high-energy boundaries composed of arrays of interfacial defects. Under intensive plastic deformation, dislocations generated within the nanocrystalline grains are swept toward the ITBs and progressively absorbed into the boundary core. The influx of abundant dislocations drives extreme local structural disorder and defect accumulation[20] at the boundaries. Supported by the high glass-forming ability of Zr[14] in the supersaturated ITBs of CuZr-1, once the localized storage of energy and free volume exceeds the crystalline lattice's stability threshold, the ITBs undergo a solid-state amorphous complexion transition. This traps the highly disordered atomic configurations at the interfaces, resulting in the deformation-induced AIF complexions at ITBs. This dislocation-dominated mechanism can also explain the superior strain hardening and ductility in CuZr-1. In contrast, CuZr-2, with low-energy CTBs and sub-threshold Zr content, lacks the structural and chemical prerequisites for such deformation-driven AIF transition.

To quantify the contributions to the strength of the films, we applied a linear superposition model comprising friction stress, Hall-Petch GB and TB hardening, and solid-solution strengthening (see Methods). Table 1 compares the model predictions with measured experimental yield strength for both alloys.

**Table 1.** Comparison of strengthening model predictions with experimental results.

| System | Strengthening model | | | | $\boldsymbol{\sigma_y}$ from Nanoindentation (MPa) | Theory-experiment difference (MPa) |
|---|---|---|---|---|---|---|
| | Friction stress $\sigma_0$ (MPa) | Hall-Petch $\Delta\sigma_{GB/TB}$ (MPa) | Solid solution $\Delta\sigma_{SS}$ (MPa) | Total $\sigma_Y$ (MPa) | | |
| CuZr-1 (primary) | 20 | 1573 | 37.6 | 1631 | 2227 | 596 |
| CuZr-2 (reference) | 20 | 1486 | 24.2 | 1530 | 1521 | 9 |

The model predicts the yield strength of CuZr-2 to within 9 MPa, validating its suitability for this alloy system. Applied to CuZr-1, the same model predicts a yield strength of 1631 MPa, which is 596 MPa much below the experimental measured value. Classical hardening mechanisms thus account for only 73% of the measured yield strength; the remaining 27% exceeds the model prediction by a factor that cannot be attributed to any mechanism captured by classical mechanisms and potentially coincides with the deformation-driven amorphous complexion transition that distinguishes CuZr-1 from the CuZr-2 (see Supplementary Note 3).

Given the observed dislocations within the deformed CuZr-1 alloy in Fig. 5, to assess whether this discrepancy could instead arise from dislocation-mediated strengthening, we applied the partial-dislocation model[40]:

$$\tau_p = \frac{2\alpha' \mu_f b_p}{d} + \frac{\gamma}{b_p} = s\,\sigma_p \quad (1)$$

where $\tau_p$ is shear stress, $\sigma_p$ denotes flow stress, $\gamma$ is the stacking fault energy, and $b_p$ represents the magnitude of the Burgers vector for a partial dislocation. The parameter $\alpha'$ denotes the character of the dislocation and is reported as 0.5[40]. Using $\mu_f$ = 42 GPa, $\gamma$ = 40 mJ m$^{-2}$, $s$ = 0.27 and $b_p$ = 0.147 nm from Cu-based systems[40,41] and $d$ = 32 nm (average grain/twin size), $\sigma_p$ is estimated as 1723 MPa for CuZr-1. This value is larger than the yield stress predicted by classical model (Table1), however it is still 504 MPa (22.6%) lower than the measured yield strength. We therefore attribute this unaccounted strengthening to the deformation-induced amorphous complexion transition and term the mechanism *amorphous complexion transformation-induced plasticity (AC-TRIP)*. Analogous to transformation-induced plasticity in steels, AC-TRIP arises from a stress-driven transformation but occurs at boundaries rather than in grain interiors. Rather than softening the host microstructure, the transformed amorphous complexions, accompanied by a possible volume change, retain high interfacial strength, consistent with the high hardness documented for CuZr metallic glasses. Because an amorphous medium lacks the translational periodicity required for crystallographic slip, abundant dislocations observed in Fig. 5 cannot propagate through the AIFs and are instead absorbed. As a result, like thermally stabilized AIFs[13,16], deformation-induced AIFs redistribute plastic strain, progressively reducing local stress concentrations and hindering crack nucleation.

To conclude, we have shown that plastic deformation alone can drive an athermal, discontinuous amorphous complexion transition at twin boundaries in a binary nanocrystalline CuZr alloy, provided the boundaries are chemically and structurally pre-conditioned. AIFs emerge at incoherent twin boundaries decorated with 9R structure under both compressive and tensile loading, persist after unloading, and form without solute redistribution—identifying deformation as a kinetic pathway for amorphous complexion transition in the supersaturated, structurally primed boundary network. Spatially resolved e-PDF provides the first comprehensive experimental evaluation of the atomic order of an amorphous complexion, revealing a continuous gradient from crystal-templated interface to glass-like core. The resulting AC-TRIP mechanism promotes plasticity and simultaneously enhances strength, ductility and toughness, contributing to over one-quarter of the yield strength. We studied CuZr as a model system, yet the required ingredients—solute supersaturation and high-energy twin boundary networks—are common to many alloy systems (see Supplementary Note 4). Our findings therefore establish mechanical deformation as a broadly applicable, non-thermal route to harness amorphous GB complexions and a design principle for damage-tolerant nanostructured metals.

## Methods

### Deposition of the CuZr films

CuZr thin films were deposited onto Si substrates at room temperature without any substrate heating using co- sputtering from pure Cu (99.999% pure) and Zr (99.2% pure) targets. Cu was placed on a radio frequency (RF) cathode due to its higher sputtering rate, while a DC cathode was used for Zr to make more precise composition control. For the CuZr-1 films, the Cu and Zr target powers were set to 80 W and 12 W, respectively, whereas the CuZr-2 films were deposited using Cu and Zr target powers of 80 W and 7 W, respectively. The deposition time was 60 min for both compositions. These deposition conditions yielded Zr concentrations of 6.2 at.% for CuZr-1 and 2.8 at.% for CuZr-2. Apart from the Zr target power, all deposition parameters were kept identical for both films. The deposited films were approximately 1 µm thick. For micropillar compression experiments, thicker films (~3 µm) with the same nominal compositions were deposited using identical sputtering conditions, with the deposition time adjusted accordingly.

### TEM-based methods (ACOM-TEM, 4D-STEM, e-PDF)

Transmission electron microscopy (TEM) was performed using a Talos microscope (Thermo Fisher Scientific) operated at an accelerating voltage of 200 kV. TEM imaging was conducted using a beam current of 3 nA, a 150 µm condenser aperture, and a spot size of 3. Image post-processing was carried out using Velox and DigitalMicrograph software. Grain size and twin thickness measurements were performed from Automated Crystal Orientation Mapping in TEM (ACOM-TEM) datasets and TEM micrographs using ImageJ, with sufficiently large representative areas analyzed to confirm statistical significance. Focused ion beam (FIB) lamellae were prepared using a dual-beam microscope. Protective platinum (Pt) depositions and a final polishing step at 2-5 kV were applied to the lamellae to minimize Ga ion beam-induced damage prior to TEM analysis[42]. The FIB lamellae extracted from both the as-deposited films and beneath the indentation imprints were approximately 20-40 nm thick.

ACOM-TEM was performed using the NanoMEGAS precession electron diffraction (PED) system and packages. ACOM-TEM and 4D-STEM data were acquired in microprobe scanning transmission electron microscopy (µP-STEM) mode using a precession angle of 0.6°, 10 precession cycles per frame, and a step size of 2.5 nm, comparable to the precession probe diameter. For ACOM-TEM maps, nanobeam electron diffraction patterns were collected at each scan position. The resulting diffraction pattern dataset was subsequently processed using the NanoMEGAS software, which indexed each pattern by matching it against a reference diffraction pattern library for FCC Cu. 4D-STEM and electron pair distribution function (e-PDF) analyses were performed using the NanoMEGAS ePDF mapping analysis software. No masking was applied during the e-PDF analysis; instead, the selected region of interest was integrated directly to obtain the S(Q) and G(r) distributions. To this end, $Q_{max}$ = 15 $Å^{-1}$, $Q_{min}$ = 0.7 $Å^{-1}$ and the composition of the material were applied.

**Nanoindentation experiments**

Nanoindentation experiments were performed using a Nano Indenter G200 (KLA). A diamond Berkovich indenter with a low tip defect was used throughout the study. Calibration on fused silica was performed prior to experiments. Prior to testing, the CuZr films on Si substrates were mounted onto flat sample holders using a thermoplastic adhesive by heating at 95 °C for 30 s.

Two types of experiments were carried out: continuous stiffness measurement (CSM) and constant stiffness relaxation measurement. A maximum indentation depth of 200 nm was applied at constant indentation strain rates (from $5 \times 10^{-3}$ $s^{-1}$ to $1 \times 10^{-1}$ $s^{-1}$) in CSM mode. Relaxation was then performed with a CSM loading at $1.5 \times 10^{-2}$ $s^{-1}$ followed by 20000 seconds holding at 200 nm. Hardness and elastic modulus were determined using the Oliver-Pharr method, with the Hay model applied to correct for substrate effects. The reported values were taken at an indentation depth corresponding to roughly 10% of the film thickness. Representative stress and strain-rate were determined following procedures proposed by Kermouche et al.[43] for CSM and Baral et al.[44] for relaxation analysis.

The yield strength of the alloys was estimated from the nanoindentation data using the method proposed by Dao et al.[45]. For each alloy, the loading slope, maximum penetration depth, unloading stiffness, and elastic modulus were extracted from the load-displacement curves obtained from 10 independent nanoindentation tests. These parameters were subsequently used as inputs to a MATLAB implementation of the dimensionless functions described by Dao et al. to estimate the yield strength[46].

**In-situ TEM Push-to-Pull (PTP) experiments**

The FIB preparation procedure for the in-situ TEM PTP tensile specimens is illustrated in Fig. S11. For the in-situ PTP tensile experiments, cross-sectional FIB chunks with approximate dimensions of 10 × 5 × 2 $\mu m^3$ were prepared and mounted onto the PTP device using an Omniprobe micromanipulator. The specimens were subsequently thinned directly on the PTP device. The central gauge section was milled, followed by removal of the Pt protective layer and the supporting substrate during the final milling step to produce freestanding CuZr tensile specimens.

In-situ TEM tensile experiments were performed using the same transmission electron microscope operated at 200 kV in bright-field imaging mode under beam-on conditions.

Mechanical loading was applied using a 100 μm conductive diamond flat-punch indenter integrated with a single-tilt PI 95 TEM PicoIndenter (Bruker Inc.). Tensile tests were conducted under displacement control at a constant loading rate of 1 nm $s^{-1}$. The applied force was obtained by correcting the raw force signal for the spring stiffness of the testing system. The specimen thicknesses, measured by high-resolution SEM, were approximately 140 nm for CuZr-1 and 135 nm for CuZr-2. The specimen width (~670 nm), total length (~2 μm), and gauge length (~1 μm) were determined from BF-TEM images. Deformation videos were recorded at 5 frames $s^{-1}$ and subsequently analyzed using in-house MATLAB scripts. Fracture toughness was evaluated by integrating the area under the engineering stress–strain curves using MATLAB.

**APT experiments**

Atom probe tomography (APT) specimens were prepared by focused ion beam (FIB) milling using a FEI Helios 600 NanoLab dual-beam instrument equipped with a micromanipulator and a gas injection system (GIS) for Pt deposition. Two distinct preparation geometries were systematically employed: top-down, enabling field evaporation perpendicular to the sample surface, and cross-section, enabling field evaporation parallel to the sample interfaces.

APT experiments were conducted on a CAMECA LEAP 6000XR instrument using laser pulsing with a deep-UV laser (λ = 257.5 nm) at a base temperature of 50 K, with a pulse energy of 30 pJ, a pulse frequency of 200 kHz, and a detection rate of 0.5%.

**Micropillar compression, SEM and XRD experiments**

In-situ micropillar compression tests were performed on thicker CuZr films (~3 μm). All pillars were milled using an FEI Helios Nanolab 660 dual beam FIB using a voltage of 30 KV, taking 3 separate steps with progressively smaller current and diameter to refine the shape. All pillars have an aspect ratio of ~2.5, with a taper angle <5°, and 5 pillars were tested for each sample to ensure repeatability. The pillars were then tested using the FemtoTools system installed inside a Gemini 360 SEM and equipped with a 5 μm diameter flat punch diamond tip. The tests were conducted at a constant strain rate of 0.05 $s^{-1}$.

The surface morphology and elemental composition of the films were examined using a Zeiss Gemini 360 scanning electron microscope (SEM) equipped with an Oxford Xplore energy-dispersive X-ray spectroscopy (EDX) system. The EDX data were acquired at an accelerating voltage of 20 kV and a working distance of 8.5 mm. In addition, the reference standard is tested prior to the tested films to ensure the accuracy of quantification.

X-ray diffraction (XRD) measurements were performed using a Bruker D8 Discover diffractometer with Cu Kα radiation (λ = 1.5406 Å) in Bragg–Brentano geometry, using an angular step size of 0.02°.

**Strengthening model**

We employed a classical linear superposition strengthening model as follows[47]:

$$\sigma_Y = \sigma_0 + \Delta\sigma_{GB/TB} + \Delta\sigma_{SS} \quad (2)$$

where $\sigma_0$ is friction stress, $\Delta\sigma_{GB/TB}$ is grain and twin boundaries strengthening (Hall-Petch), and $\Delta\sigma_{SS}$ is solid solution strengthening contribution. $\Delta\sigma_{GB/TB}$ can be written as following[48]:

$$\Delta\sigma_{GB/TB} = k_{GB} d^{-1/2} + k_{TB} \lambda^{-1/2} \quad (3)$$

where $k_{GB} = k_{TB}$ is reported as 3478 MPa nm$^{1/2}$ from[48]. Also, $d$ and $\lambda$ are grain size and twin size, respectively. Substituting the measured grain sizes of 54 nm and 63 nm, and twin thicknesses of 10 nm and 11 nm, for CuZr-1 and CuZr-2, respectively, yields $\Delta\sigma_{GB/TB} = 1486\ MPa$ for CuZr-2 and $\Delta\sigma_{GB/TB} = 1573$ for CuZr-1.

$\Delta\sigma_{SS}$ can be written as following[47]:

$$\Delta\sigma_{SS} = \Delta\sigma_{Fleischer} + \Delta\sigma_{nc} \quad (4)$$

where $\Delta\sigma_{Fleischer}$ represents the classical solid-solution strengthening contribution, while $\Delta\sigma_{nc}$ proposed by Rupert et al.[49] accounts for the additional strengthening arising from solute-induced pinning of dislocations by grain boundaries in nanocrystalline materials, referred to as nanocrystalline solution-pinning strengthening.

$\Delta\sigma_{Fleischer}$ can be written as follows[47,49]:

$$\Delta\sigma_{Fleischer} = A\ G_{solvent}\ \varepsilon_S^{3/2}\ c^{1/2} \quad (5)$$

Assuming the constant ($A$=0.126) from[47] and $c$ (nominal solute content at.%) of 2.8 and 6.2 for CuZr-2 and CuZr-1, respectively, the Fleischer strengthening contribution can be estimated. In $\varepsilon_S = |\varepsilon'_G - m\varepsilon_b|$, (m=3) from[47], the terms $\varepsilon_b$ and $\varepsilon'_G$ denote the lattice mismatch coefficient and the modulus mismatch coefficient, respectively, which are defined as $\varepsilon_b = (\frac{db}{dc})/b$ and $\varepsilon'_G = \frac{\varepsilon_G}{1+0.5\varepsilon_G}$, where $\varepsilon_G = (\frac{dG}{dc})/G$. The terms $\frac{db}{dc}$ and $\frac{dG}{dc}$ for CuZr films are found to be 0.0185 nm and 86.1 GPa from literature[47]. Now by normalizing to $b$ (interplanar spacing of pure Cu, i.e. 0.0208) and $G$ (shear modulus 48 GPa) to get $\varepsilon_G$ and $\varepsilon_G$, and by substituting to above equations, giving $\varepsilon'_G$=0.945 and $\varepsilon_b$=0.889. And therefore $\varepsilon_S = 1.722$ is attained. Finally, by substituting these values in Fleischer equation ($Eq. 5$) the following results yield: $\Delta\sigma_{Fleischer} = 22.8$ MPa, and $\Delta\sigma_{Fleischer} = 34$ MPa for CuZr-2 and CuZr-1, respectively.

Now we consider $\Delta\sigma_{nc}$ proposed by Rupert et al.[49]:

$$\Delta\sigma_{nc} = \frac{G_{solvent} b_{solvent}}{d}\ \varepsilon_{nc}\ c \quad (6)$$

where $\varepsilon_{nc} = \frac{dG/dc}{G} + \frac{db/dc}{b}$ which equals to 2.68 by substituting the values discussed above. Where $d$ (average grain size) is 37 nm and 32 nm for CuZr-2 and CuZr-1, respectively. Here, the reported grain size represents the average of both the columnar grain width and the twin thickness, as presented previously. $b$ denotes the Burgers vector magnitude of the Shockley partial dislocation and is ~0.1479 nm[41]. Applying $G$ (shear modulus) and $c$ (nominal solute content at.%) for each alloy like above, $\Delta\sigma_{nc}$ amounts to 1.43 MPa and 3.67 MPa for CuZr-2 and CuZr-1, respectively. So, the total $\Delta\sigma_{SS}$ contribution by ($Eq. 4$) is 24.23 MPa and 37.67 MPa for CuZr-2 and CuZr-1. The overall $\sigma_Y$ from ($Eq. 2$) equals to 1530 MPa and 1631MPa for CuZr-2 and CuZr-1, respectively (considering $\sigma_0 = 20$ MPa[50]). The predicted results of the strengthening model are presented in Table 1.

## Authors contributions

Masoud Ahmadi (M.A.) conceived the idea, designed the research approach, and directed the project. M.A. performed the TEM imaging, ACOM-TEM, 4D-STEM, in-situ TEM PTP, nanoindentation, and e-PDF experiments, analyzed the resulting data, prepared the figures and wrote the original manuscript. Jin Qin (J.Q.) deposited the films and performed the SEM, XRD, and micropillar compression experiments and corresponding data analysis. Gabrielle Tiphene (G.T.) conducted the nanoindentation and relaxation experiments and analyzed the corresponding data. Mohamed Charai (M.C.) prepared the APT specimens and performed the APT experiments/analysis. Alejandro Gomez Perez (A.G.P.) contributed to ACOM-TEM and e-PDF techniques and corresponding discussions. Khalid Hoummada (K.H.) analyzed and interpreted the APT results. Thomas Pardoen (T.P.) contributed to the interpretation and discussion of the nanomechanical testing results and provided resources. Matteo Ghidelli (M.G.) supervised J.Q. during the film deposition and characterization and contributed to the discussion of the results and provided research resources. Hosni Idrissi (H.I.) contributed to the scientific discussions, provided guidance on the interpretation of the results, and provided research resources. All authors read and commented on the manuscript and contributed to its revision.

## Acknowledgments

Masoud Ahmadi acknowledges the LACAMI platform and Armand Béché at iMMC/UCLouvain for the facilities and technical support. Hosni Idrissi is mandated by the Belgian National Fund for Scientific Research (FSR-FNRS). Matteo Ghidelli and Jin Qin acknowledge the financial support of the French National Research Agency (ANR) projects EGlass (N° ANR-22-CE92-0026-01) and Super-Glasses (N°. ANR-24-CE91-0002-02).